# Ordering stakeholder viewpoint concerns for holistic and incremental Enterprise Architecture: the W6H framework

MUJAHID SULTAN, Government of Ontario, Ryerson University and

ANDRIY MIRANSKYY, Ryerson University

**Context**: Enterprise Architecture (EA) is a discipline which has evolved to structure the business and its alignment with the IT systems. One of the popular enterprise architecture frameworks is Zachman framework. This framework focuses on describing the enterprise from six viewpoint perspectives of the stakeholders. These six perspectives are based on English language interrogatives 'what', 'where', 'who', 'when', 'why', and 'how' (thus the term W5H). Journalists and police investigators use the W5H to describe an event or incident. However, it is difficult to "fit" EA into the "universe" of events, leading to difficulties in creation and evolution of EA. Moreover, the ordering in which interrogatives should be answered and viewpoints arranged is not defined in the existing EA frameworks, making it difficult to decide which of the viewpoint concerns should be captured first?

Our **goals** are to 1) assess if W5H is sufficient to describe EA of today's rapidly evolving enterprise, and 2) explore the ordering and precedence among the viewpoint concerns based on interrogative questions.

**Method**: we achieve our goals by bringing tools from the Linguistics, focusing on a full set of English Language interrogatives to describe viewpoint concerns and the inter-relationships and dependencies among these. Application of these tools is validated using pedagogical EA examples.

**Results**: 1) We show that addition of the seventh interrogative 'which' to the W5H set (we denote this extended set as W6H) yields extra and necessary information enabling creation of holistic EA. 2) We discover that particular ordering of the interrogatives, established by linguists (based on semantic and lexical analysis of English language interrogatives), define starting points and the order in which viewpoints should be arranged for creating complete EA. 3) We prove that adopting W6H enables creation of EA for iterative and agile SDLCs, like Scrum.

**Conclusions**: We believe that our findings complete creation of EA using Zachman framework by practitioners, and provide theoreticians with tools needed to improve other EA frameworks such as TOGAF and DoDAF.

## 1. INTRODUCTION

Enterprise Architecture frameworks consist of a set of artefacts which are description of the enterprise from specific viewpoint[1] of a group of stakeholders (Finkelstein et al. 1992; Rozanski and Woods 2011). The stakeholders are generally grouped as owners, designers (architects), systems engineers and developers. A number of frameworks have evolved over time (Pereira and Sousa 2004; Sessions 2007) and have received some maturity (Shah and Kourdi 2007) over the past decade, TOGAF, FEAF, Zachman to name a few.

John Zachman introduced the concept of Information System Architecture (ISA) in 1987 (Zachman 1987). The Zachman framework describes stakeholders' views focusing on five Wh-interrogatives ('what', 'who', 'wher'e, 'why', and 'when') and one H-Interrogative ('how'). This focus comes from journalism's W5H theory (Flint 1917). Zachman framework (given in Table 1) consists of two dimensions: views of a particular stakeholder group of the enterprise from a particular perspective (rows of Table 1) and the description of these views (cells of Table 1). The description information is gathered by answering six out of the seven English language interrogatives ('what', 'how', 'where', 'who', 'when', and 'why'). Zachman argued (Sowa and Zachman 1992; Zachman 1999; Zachman 1987) that answering these interrogatives from the view point of Owner, Designer, Builder and Sub-Contractor enables the development of Information System Architecture (ISA). Note that Table I represents a grid of somehow associated but mostly disjoint artefacts (John Zachman 2002). In this publication we propose a link among the columns based on semantic and lexical rules if English language interrogatives and identify the missing interrogative in the table I and its importance in creation of complete EA.

---

[1] We use ANSI/IEEE Standard 1471-2000 definitions of stakeholder views and viewpoints (Kruijff et al. 2007). A view is a representation of a whole system from the perspective of a related set of concerns. A viewpoint defines the perspective from which a view is taken. In other words, a viewpoint is where you are looking from - the vantage point or perspective that determines what you see; a view is what you see.

Table I: Zachman Framework of Enterprise Architecture

|  | Data (What) | Function (How) | Network (Where) | People (Who) | Time (When) | Motivation (Why) |
|---|---|---|---|---|---|---|
| **Scope (Ballpark View)** | List of things important to business | List of processes the business performs | List of location in which the business operates | List of organizations important to the business | List of events/cycles significant to the business | List of business goals /strategies |
| **Business Model (Owners' View)** | e.g. Semantic Model | e.g. Business process model | e.g. Business logistics Systems | e.g. Workflow model | e.g. Master schedule | e.g. Business Plan |
| **System Model (Designer's View)** | e.g. Logical data Model | e.g. Applications architecture | e.g. distributed Systems architecture | e.g. Human interface architecture | e.g. Processing structure | e.g. Business Rule Model |
| **Technology Model (Builder's View)** | e.g. Physical data Model | e.g. System design | e.g. Technology Architecture | e.g. Presentation architecture | e.g. Control structure | e.g. Rule design |
| **Detailed Representations (Subcontractor)** | e.g. Data Definition | e.g. Program | e.g. Network architecture | e.g. Security architecture | e.g. timing definitions | e.g. rule specification |
| **(Functioning System)** | e.g. Data | e.g. Function | e.g. Network | e.g. Organization | e.g. Schedule | e.g. Strategy |

The rest of this paper is structured as follows. In Section 2, we discuss the basic set of English language interrogatives, interdependencies among these and their relation to viewpoint description. Based on these findings, we propose a framework for effective EA in Section 3. We discuss application of the framework in Section 4. In Section 5 we discuss the proposed framework. Finally, Section 6 presents the conclusions and future work.

## 2. MAJOR ISSUES WITH APPLICATION OF THE W5H TO DEFINE ENTERPRISE ARCHITECTURE

The application of W5H to articulate stakeholder views of an enterprise raises the following obvious questions (hence our research questions):

**RQ1**: is the W5H set of interrogatives complete? Is there any interrogative missing?
**RQ2**: can six interrogatives in W5H set describe stakeholder viewpoints for holistic EA?
**RQ3**: which of these six interrogatives should be asked first? And is there any inter-dependency?

**RQ4**: is the order in which these interrogatives are asked (viewpoints arranged) for EA important? Is there any such order?

**RQ5**: can W5H-based EA frameworks handle iterative and agile developments?

To answer these questions we look at the enterprise architecture from:
(a) Lexical and semantics of Wh-interrogative questions perspective, to see if any interrogative is missing or is the order of interrogatives important;
(b) Stakeholder viewpoint perspective to establish why W5H is not sufficient, to capture the EA;
(c) SDLC perspective, to show why W5H is insufficient for iterative and agile SDLCs.

### 2.1 Semantics and syntax of English Language interrogative questions (inter-relationships and dependencies among these)

The description of stakeholder viewpoints by asking six (W5H) interrogative questions as described by (Finkelstein et al. 1992; Rozanski and Woods 2011; Sowa and Zachman 1992; Zachman 1987) does not cater to the order in which these interrogatives should be asked (John Zachman 2002). However, in English language, these interrogatives have a precedence relationship.

Ginzburg (Ginzburg and Sag 2000) illustrated the fecundity of Head-driven Phrase Structure Grammar (HPSG) as a framework. It is regarded as the most explicit description of syntax and semantics of English Language Interrogatives of any era of English syntax (Koenig 2004). Cysouw further elaborated Ginzburg's work by presenting the morphological and lexical analysis of English Language Interrogatives in his papers ( Cysouw 2005) and described the precedence relationship among the interrogatives as shown in Figure 1. Notice the interrogative 'which', missing from the W5H, playing an important role in the precedence relationship among the interrogatives.

In the majority of the world languages 'who', 'what', 'which' and 'where' are four basic lexemes, referred to as the 'major four' (Lindström 1995). Cysouw (Cysouw 2005) further described the typology of interrogative categories as follows:

• The major categories: person (who), thing (what), selection (which), place (where);
• The minor categories: quantity, manner (how), time (when);
• The incidental categories: reason (why), quality, extent, position, action, rank, etc.

The discussion above shows that a very important interrogative 'which' is missing in W5H-based frameworks and the order in which these interrogative questions should be asked is important to satisfy the needs of next interrogative question. Thus, applying W5H to capture viewpoint concerns of stakeholders is not sufficient, notice the missing 'which' interrogative and precedence relationship among the interrogatives. It is evident that without 'which' interrogative we cannot answer the remaining dependent interrogatives properly, as shown in Figure 1, and capturing and organizing stakeholder viewpoint concerns without an order leads to information loss.

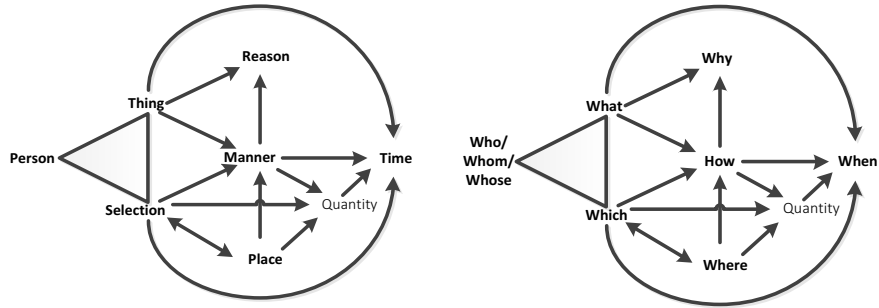

Fig. 1. Right: Order and inter-dependencies of English Language Interrogatives based on (Cysouw 2004) and (Koenig 2004). Left: material categories corresponding to the interrogative words. Legend: The edges represent relationships among interrogatives, and the arrows point to dependent interrogatives. Directionless edges indicate an absence of dependency among interrogatives. Bidirectional arrows indicate interrogatives are interchangeable and have stong dependency among each other.

Some research has been conducted (Pereira and Sousa 2004) to establish a starting point (or a starting viewpoint) for the Zachman framework of EA, but no concrete evidence has been provided, which can establish such a starting point or propose an order. We show that there is an order in the capture and description of the stakeholder viewpoints and we propose a starting point in Section 3. To the best of our knowledge, no research has been done to establish the impact of precedence relationship among the English language interrogatives on describing Enterprise Architecture from stakeholder viewpoints perspective. However in our recent publication we have applied the order of interrogative questions in requirements engineering domain (Mujahid Sultan and Andriy Miranskyy 2015).

It is tempting to substitute 'which' with 'what' or 'who', as often happens during informal conversations. However, in the formal settings, this would lead to loss of information. Wh-phrases do not introduce quantifiers into semantic representations but rather parameters (Ginzburg and Sag 2000); Wh-phrases always scope wider than Generalized Quantifiers (Koenig 2004). Koeing showed that multiple Wh-questions such as Q1 and Q3 are not semantically equivalent to quantified Wh-questions as Q2:

Q1: "Who proved what?"
Q2: "Who proved each theorem?"
Moreover, Wh-phrases act as quantifiers as shown in the question Q3:
Q3: "Which student read which book for which course?"
They further show that 'which' cannot be substituted with 'what', as shown in Q4 and Q5. 'Which' quantifies the selection, whereas 'what' is infinite:
Q4: Which requirements the stakeholder group liked the most?
Q5: What requirements the stakeholders liked the most?

This illustrates the inter-relationships and dependencies among interrogative questions and establishes the importance and necessity of the 'which' interrogative in answering the dependent interrogative questions. We discuss our proposed solution to this problem in Section 3.

### 2.2 Enterprise is not an Event

W5H theory has its roots and utilization in the field of journalism (Flint 1917; Griffin 1949). The six interrogatives can describe an incidence from a journalist's point of view: what happened, when it happened, who was involved, where it happened, and why it happened. Application of W5H theory to describe an enterprise as suggested by Zachman framework for EA seems insufficient, as an enterprise is not an event.

The description of an enterprise from stakeholder's viewpoint is not an incident; rather it is a description of either an existing or functioning enterprise, or the description of a new enterprise to be created. Using W5H for the description of the enterprise does not cater for the complete needs of stakeholders unless we answer the seventh Wh-Interrogative 'which'. From an incident point of view, answering what happened and where might be enough to describe an incident: for example, the news 'five people were killed in an accident at the bridge' uses W5H without 'which' and gives "good enough" picture[2]. Whereas, if we are trying to describe the enterprise from the viewpoints of the stakeholders (or stakeholders' groups), we need to know which stakeholders need what. Skipping 'which' will not capture the whole picture.

Furthermore, enterprise is not a static entity and undergoes continuous changes: new products and services are offered periodically, and business and process mergers occur all the times. This warrants the designers and business owners to plan for new products and services and strategically position these to be provisioned in the future dates. W5H does not have any mechanisms for such kind of strategic planning as all the interrogatives are incidence based and deal with a point in time. We discuss our proposed solution to this problem in Section 5.

### 2.3 Iterative System Development Lifecycles

Based on the authors' industrial experience, W5H-based EA frameworks, such as Zachman framework, work only for Waterfall Software Development Life Cycle (SDLC) and do not provide any mechanisms for modern SDLC approaches like Agile and Iterative. Create, Read, Update and Delete (CRUD) matrix has gained most popularity in object oriented analysis and design (OOAD) (Daniel Brandon Jr. 2002). CRUD matrix plays an important role in describing the link between data entities to the software components (classes, components, modules, etc.). Unfortunately, W5H-based frameworks (e.g., Zachman Framework) cannot define CRUD matrix due to the missing 'which' interrogative, hence leaving the gap for Agile and Iterative SDLCs that require selection.

Some modern approaches for rapid software development, e.g., Scrum, heavily depend upon selection ('which'). In Scrum the product manager "selects" features from the 'product backlog' to be included in a 'sprint'. Similarly, the 'scrum master' selects 'sprints' to be part of a 'release'. As there is no selection mechanism in the W5H-based frameworks, Scrum cannot leverage these frameworks. We further discuss our proposed solution for this problem in Section 5.

## 3. SOLUTION: W6H TO CAPTURE AND DEFINE HOLISTIC ENTERPRISE ARCHITECTURE

### 3.1 The importance of 'which' interrogative to describe stakeholder view / define EA

In the following we analyze the W6H and qualify the 'which' interrogative from stakeholder groups' perspective. We take the Zachman's definition of Stakeholder views, described below:

- Scope (Ballpark View),
- Business Model (Owner's View),
- System Model (Designer's View),
- Technology Model (Builder's View),
- Detailed Representations (Subcontractor's View),
- (Functioning System).

We build our argument by analyzing the first four of the six stakeholder views (namely ballpark, owner's, designer's, and builder's views) given in the first column of the Table I. The remaining two views (subcontractor and the functioning system views) are outsourced to external vendors (Sowa and Zachman 1992); therefore, we can ignore these two views, without loss of generality.

---

[2] As the reporter typically does not care how many people were passing by at that time and 'which' ones got killed, s/he is only concerned with who were killed.

3.1.1 Scope (Ballpark View): Ballpark view sets the scope and puts architecture effort in perspective, and is also called the 'contextual' view. Second row of the Table I presents the W5H perspective of the contextual view. This view focuses on a list of things, functions, locations, organizations, events, goals, and strategies important to the business. Asking and adding the 'which' interrogative to this view provides a new perspective and adds to the context by enabling selection among these lists. This also makes these lists finite[3], allowing one to choose and select.

3.1.2 Business Model (Owner's View): The third row of Table I, 'Business Model' describes owner's view of the enterprise. 'What the enterprise does and why' is captured in the 'business plan'. A business plan describes the value proposition and customer segmentation for the business. "For whom are we creating the value" and "which ones of our customers' problems we are helping to solve" are the usual value propositions. Consider the following questions for service delivery channels to customers:

- Which customer needs are we satisfying?
- Through which channels do our customer segments want to be reached?
- How are we reaching them now?
- How are our channels integrated?
- Which ones work best?
- Which ones are most cost-effective?

Consider the following questions for partners and suppliers:

- Who are our key partners and suppliers?

- Which resources are we acquiring from partners?
- Which key activities do partners perform?

It is evident from these questions/discussion that description of value position/service delivery channels requires seventh Wh-interrogative 'which' to complete the 'Business Model' or the owners' view of the enterprise.

3.1.3 System Model (Designer's View): Consider the fourth row of the Table I, the 'System Model'. Systems model is the view of enterprise from system designers (automation perspective), where the business processes (candidates for automation) are described in systems terms. At this stage the designers (system engineers/architects) identify business concepts (entities) on which the system components will work (use-cases, application components, etc.), or the entities used by system functions. As mentioned in Section 2.3, the CRUD matrix is used in OOAD to identify these business concepts (entities).

W5H-based frameworks do not provide any mechanism to create CRUD matrix unless we introduce the 'which' interrogative question (the selection). 'Which' data entities are used by which application functions? We can argue that the entities and the business/application functions can be defined and developed separately but it is the 'which' interrogative that enables the link between them.

3.1.4 Technology Model (Builder's View): Technology view is the description of systems/enterprise from the technology and infrastructure perspective, the fifth row of Table I. This information, for example, aids in planning for disasters. Business Continuity Planning (BCP) and Disaster recovery planning (DRP), part of business continuity Management (BCM), deal with service disruption of an organization's operations. DRP involves a) identification of the critical business functions that are essential to the continuity of the enterprise's business and b) identification of resources that are key to the operations of these business functions, e.g., data, people, and locations.

---

[3] Implicitly, these lists are infinite, as discussed in Section 2.1.

The first step in BCP/DRP involves 'selection' of business functions critical to the business among a finite set (after assessing the value of all functions) thus proving the need and necessity of the 'which' interrogative for BCP and DRP. A typical question to ask in DRP is: which systems are more critical to be recovered first? This enables to plan for disaster recovery site and type.

### 3.2 The Order of the Interrogatives

In the following we discuss some cases of dependence among the interrogatives from business and systems perspective (following discussion is based on Figure 1); we summarize interrelations between interrogatives in Table II.

3.2.1 Case 1 (to answer 'how' interrogative, we need to answer 'what' or 'which'):    In Figure 2 we highlight examples of the precedence and dependency relationship among the English language interrogatives. For example, in the Figure 2a, to answer the 'how' interrogative question correctly one needs to know either 'what' or 'which'

Consider the following sentences:

- How beautiful it is!
- How inefficiently they run this place!
- How badly the brown car is broken!

These sentences show that 'how' is dependent on information that can be obtained using either 'what' or 'which' questions. Without this information the sentences above will read as follows:
- How beautiful ~~it~~ is!
- How inefficiently they run ~~this place~~!
- How badly ~~the brown car~~ is broken!

This is very much true in the systems domain; consider Figure 2b: to know the 'functions' of a software system one needs to know the 'data' the function will work on[4].

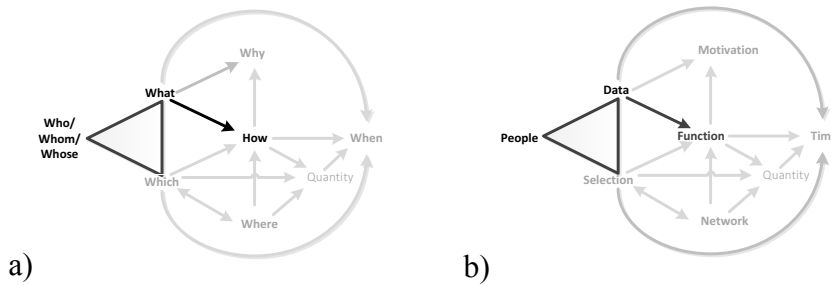

Fig. 2.    (a) Dependency of 'how' on 'what' interrogative question (b) Dependency of System functions on data

3.2.2 Case 2 (to answer 'why', we need to know 'how' or 'what'): The following sentences explain the dependency of 'why' interrogative on 'how' and 'what' (the underlined words).

- Why did Carrie cross the road?
- Why has there never been a President of the United States named Clovis?
- We have been told why he is writing this paper!
- Why is the train delayed?
- I wonder why (this includes how) the boy was injured?

Similarly in systems terms, in Figure 3b the highlighted sections are interpreted as follows: "the organizations/people/actors (and their goals thereof) interested in the data (things) to perform (a list) of processes 'when' and 'why'". This dependency makes very logical sense in software architecture: one needs to select the functions that work on some data, or the data elements which are used by some function. Without this selection, very important component of the software architecture, the relationship among the data entities and functions cannot be captured.

---

[4] Not all software systems are data-driven; however, the majority of business systems are.

Similarly, to describe the 'owners view' in EA (which is first row in Zachman Framework, second row of Table I) the starting point (interrogative) is 'who/whom/whose' (the people), and the next interrogative is 'what' (data) they are interested in, leading to the description of the next interrogative 'how' (function), meaning what functions needs to be described using this data to satisfy the needs of these people. From this point on we can describe 'when' (the business cycles) and 'why' (business motivation) quite easily.

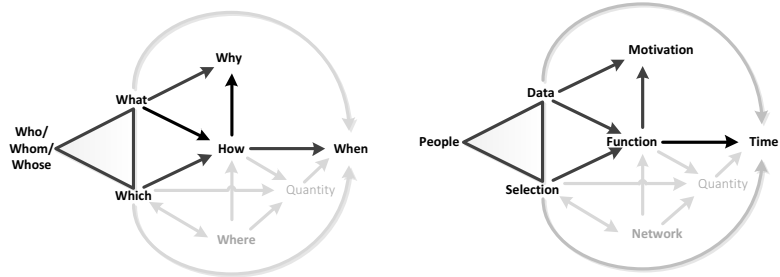

Fig. 3. Dependency of 'why' on 'what' and 'how' interrogative questions (b) Dependency of System Functions on selection (of data elements)

3.2.3 Case 3: From the software architecture perspective it will be difficult to describe which requirements are fulfilled by which application, and which application is deployed on which system/network, and satisfy the needs of which geographic locations if we don't answer the 'which' interrogative as shown in the Figure 4.

Similarly, from the enterprise architecture perspective, it will be difficult to describe the locations where business operates, unless we answer the 'which' interrogative question.

In order to answer 'why', we need to know 'how' or 'what', and 'how' is further dependent upon 'where'. Only the description of 'which' interrogative question can justify 'why' a business function is provided at a specific location. On Figure 4 follow the arrow from 'where' to 'how' to 'why'. To answer 'where' interrogative question we need to do the selection first, meaning that 'where' interrogative question has a strong dependency on the selection, the 'which' interrogative question.

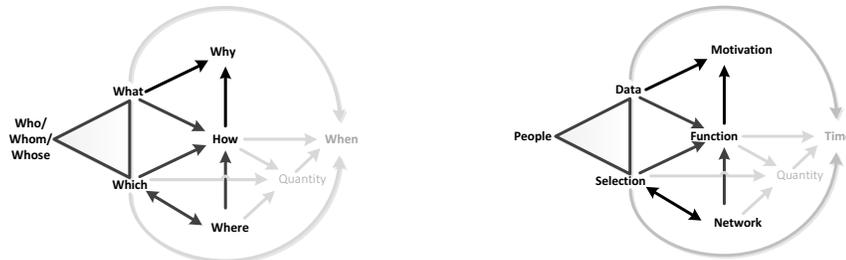

Fig. 4. (a) Dependency of 'how' on 'what' interrogative question (b) Dependency of System Functions on Data

**3.3** The importance of 'which' interrogative for Iterative and Agile development

A very important task of agile and iterative developments is to identify data required by system functions. Create, Read, Update and Delete (CRUD) matrices became prevalent with emergence of object oriented analysis and design (Daniel Brandon Jr. 2002). CRUD matrices provide an easy mechanism to associate and link system functions with data elements. The 'which interrogative question' is key in creating CRUD matrix. W5H-based approaches cannot create a CRUD matrix unless which interrogative question is introduced for the purposes of selection. In other words, we need to establish which data entities are to be used by which application functions. Although we might elicit requirements for entities and application functions separately, it is the 'which' interrogative that establishes a link between data and functions.

Table II. The proposed W6H Framework for defining holistic Enterprise Architecture. Cells with number in round brackets can be defined in any order; dependency of a given cell X on cells Y and Z is recorded using "X → Y, Z" notation.

|  | People (Who) (1) | Data (What) (2) | Selection (Which) (3) | Network (Where) (4) | Function (How) (5 →2,3 or4) | Motivation (Why) (6→2,5) | Time (When) (7→5,6) |
|---|---|---|---|---|---|---|---|
| **Scope (Ballpark View)** | List of organizations important to the business | List of things important to business | Which things, business processes, clients / organizations, goals and strategies are important to the business | List of location in which the business operates | List of processes the business performs | List of business goals /strategies | List of events/cycles significant to the business |
| **Business Model (Owners' View)** | e.g. Workflow model | e.g. Semantic Model | e.g. <br>- Which business entities<br>- Which business processes<br>- Which business scenarios<br>- Which business functions<br>- Which clients / organizations | e.g. Business logistics Systems | e.g. Business process model | e.g. Business Plan | e.g. Master schedule |
| **System Model (Designer's View)** | e.g. Human interface architecture | e.g. Logical data Model | e.g. <br>- Which logical entities<br>- Which system use cases<br>- Which system functions<br>- Which applications<br>- Which data centers<br>- Which business rules | e.g. distributed Systems architecture | e.g. Applications architecture | e.g. Business Rule Model | e.g. Processing structure |
| **Technology Model (Builder's View)** | e.g. Presentation architecture | e.g. Physical data Model | e.g. <br>- Which physical entities<br>- Which application components<br>- Which technology components | e.g. Technology Architecture | e.g. System design | e.g. Rule design | e.g. Control structure |
| **Detailed Representations (Subcontractor)** | e.g. Security architecture | e.g. Data Definition | e.g. <br>- Which data instance<br>- Which program component<br>- Which network segment | e.g. Network architecture | e.g. Program | e.g. rule specification | e.g. timing definitions |
| **(Functioning System)** | e.g. Organization | e.g. Data | e.g. <br>- Which Data<br>- Which Function<br>- Which Network | e.g. Network | e.g. Function | e.g. Strategy | e.g. Schedule |

In our experience, practitioners do use 'which' interrogative question to link entities and system functions and to create CRUD matrices, but, due to lack of any formal methodology, it is not practiced consistently. The proposed framework, given in Section 4, provides selection mechanism needed for iterative and agile SDLCs like Scrum.

## 4. PROPOSED FRAMEWORK

We have shown that without answering the seventh English Language interrogative, 'which' we cannot describe the dependent interrogatives and hence cannot capture stakeholder perspectives completely. We also showed that the order of the interrogatives (order in which stakeholder viewpoint concerns are arranged) plays a major role in the description of interrogatives/viewpoint concerns. Based on these findings, we propose W6H framework for describing the enterprise architecture of today's evolving enterprise.

We propose that 'who', 'what', 'which' and 'where' interrogatives can independently exist and are not dependent upon any interrogative. Note that the double sided arrow between 'which' and 'where' indicates that these two interrogatives can be interchanged. To answer 'how' interrogative properly, we must answer the 'what' interrogative and any of the 'which' or 'where' interrogatives. The 'why' interrogative can only be answered if 'what' and 'how' interrogatives are already answered. 'When' interrogative is dependent upon 'how' and 'why' interrogatives. Table II presents the proposed model with additional 'which' perspective and prescribes an order in the description of the stakeholder perspectives.

## 5. DISCUSSION

### 5.1 Starting Point and Order:

EA community using the Open Group's Architecture Development Method (ADM) and Zachman framework frequently struggle to find a starting point and order in which EA can be captured (Pereira and Sousa 2004). John Zachman (John Zachman 2002) argues that there is no such starting point as all the interrogatives are independent, but we have shown that this is not the case. There is an order in the interrogatives and some interrogatives are dependent upon others as described in Section 3.

### 5.2 Incremental and TO BE views:

US federal statute (Cohen 1996) require public enterprises to submit their Enterprise Architecture to the government. These EA submissions must provide strategies that enable the agency to support its current state and transition to its target state. We showed that a very important benefit of asking the seventh English language interrogative 'which' is that it enables creation of the incremental view of the enterprise from stakeholder viewpoints, as shown in the Table III. This enables enterprise architects to create historical views of the enterprise (why the enterprise went through a specific path in history) enabling creation of AS-IS[5] and TO-BE[6] views of the architecture.

Referring back to section 3.1.1, this also enables the owner to describe contextual view of the enterprise in increments (to align with strategic goals). Thus enabling defining "AS-IS" and "TO-BE" views at the contextual level.

### 5.3 Agile and Iterative Development:

We also showed that most prevalent EA frameworks are based on the waterfall system development life cycles and do not have any mechanism of dealing with the modern iterative and agile methodologies. The major benefit of introduction of the 'which' interrogative in W5H-based frameworks is that it enables the selection mechanism, enabling creation of Enterprise Architecture for iterative and agile methodologies like Scrum.

---

[5] AS-IS: refers to the current state (or the description of the current state) as is, without future vision or plans.
[6] TO-BE: refers to the desired/planned or future state (or the description of future/desired/planned state).

Table III. The proposed W6H Framework for defining incremental Enterprise Architecture

| | People (Who) (1) | Data (What) (2) | Selection (Which) (3) | Function (How) (4) | Network (Where) (5 → 2,3/4) | Motivation (Why) (6 → 2,5) | Time (When) (7 → 5,6) |
|---|---|---|---|---|---|---|---|
| **Scope (Ballpark View)** | List of organizations important to the business | List of things important to business | Which things, business processes, clients / organizations goals and strategies are important to the business | List of processes the business performs | List of location in which the business operates | List of business goals /strategies | List of events/cycles significant to the business |
| **Business Model (Owners' View)** | e.g. Workflow model | e.g. Semantic Model | e.g. - Which business entities - Which business processes - Which business scenarios - Which business functions - Which clients / organizations | e.g. Business process model | e.g. Business logistics Systems | e.g. Business Plan | e.g. Master schedule |
| **System Model (Designer's View)** | e.g. Human interface architecture | e.g. Logical data Model | e.g. - Which logical entities - Which system use cases - Which system functions - Which applications - Which data centers - Which business rules | e.g. Applications architecture | e.g. distributed Systems architecture | e.g. Business Rule Model | e.g. Processing structure |
| **Technology Model (Builder's View)** | e.g. Presentation architecture | e.g. Physical data Model | e.g. - Which physical entities - Which application components - Which technology components | e.g. System design | e.g. Technology Architecture | e.g. Rule design | e.g. Control structure |
| **Detailed Representations (Subcontractor)** | e.g. Security architecture | e.g. Data Definition | e.g. - Which data instance - Which program component - Which network segment | e.g. Program | e.g. Network architecture | e.g. rule specification | e.g. timing definitions |
| **(Functioning System)** | e.g. Organization | e.g. Data | e.g. - Which Data - Which Function - Which Network | e.g. Function | e.g. Network | e.g. Strategy | e.g. Schedule |

## 6. SUMMARY AND FUTURE WORK

In this paper, we sought to answer the following research questions to effectively define/describe the enterprise architecture.

**RQ1**: is the W5H set of interrogatives complete? Is there any interrogative missing?
**RQ2**: can six interrogatives in W5H set describe stakeholder viewpoints for holistic EA?
**RQ3**: which of these six interrogatives should be asked first? And is there any inter-dependency?
**RQ4**: is the order in which these interrogatives are asked (viewpoints arranged) for EA important? Is there any such order?
**RQ5**: can W5H-based EA frameworks handle iterative and agile developments?

We applied linguistic findings to answer **RQ1**, and demonstrated that English language has set of seven basic interrogatives and demonstrated that W5H-based frameworks are missing a very important interrogative 'which'. We showed that extending the set of interrogatives 'what', 'where', 'when', 'who', 'why', and 'how' (denoted as W5H) with interrogative 'which' (we denote it as W6H) enables creation of holistic EA, answering **RQ2**. We showed that asking questions, based on the W6H (in the order of precedence described in Figure 1 and Table II) improves information flow for description of the stakeholder viewpoints, answering **RQ3** and **RQ4**. We also discussed that the 'which' interrogative makes selection and prioritisation possible, enabling creation of EA for iterative SDLCs, answering **RQ5**.

Finally, we created Table II, based on the W6H framework, to capture stakeholder viewpoint concerns for holistic EA. Table II serves as an effective tool for creating and capturing EA for iterative and agile SDLCs. We then extended the framework for iterative EA, given in Table III, which enables one to create 1) a blueprint of the enterprise at a specific point in time and 2) a plan for the next stage. This links "snapshots" of the entire organization at various points in time, enabling creation of incremental EA.

We presented use cases of the W6H framework form every stakeholder groups' viewpoint perspective and demonstrated that the framework guarantees completeness in capturing stakeholders' viewpoint concerns, enabling creation of holistic EA. We showed that not following the framework might leave gaps in stakeholder viewpoint concerns, leading to incomplete EA. We also stressed that the missing 'which' interrogative in W5H-based patterns plays an important role in the selection process and provides effective mechanisms for iterative and agile developments.

We believe that our findings are of interest to practitioners who can readily use the W6H framework to capture complete set of viewpoint concerns of stakeholders and create a holistic EA of their enterprise. The findings will also be of the interest to theoreticians. The linguistic theories provide a sound foundation for the extension and generalisation of our framework, enabling novel work in stakeholder viewpoint capture and description. These findings are based on the authors' two decades of experience in numerous large-scale, complex enterprise projects in the public and private sectors. In this paper, we gave pedagogical examples of the W6H framework usage based on real-world use cases frequently encountered by the authors. We plan to formally validate the pattern using datasets collected from industrial projects.

Disclaimer: The opinions expressed in this paper are those of the authors and not necessarily of the Government of Ontario.